\documentclass[aps,twocolumn,preprintnumbers,amsmath,amssymb,nofootinbib,superscriptaddress,notitlepage]{revtex4-1}

\usepackage{slashed}
\usepackage{soul}
\usepackage{mathrsfs,bm}
\usepackage{txfonts}
\usepackage{amssymb}
\usepackage{indentfirst}
\usepackage{graphicx,booktabs}
\usepackage{multirow}
\usepackage{overpic}
\usepackage{color}
\usepackage{amssymb}
\usepackage{epsfig}

\usepackage[utf8]{inputenc}

\begin{document}
\title{The $P_{\psi s}^{\Lambda}(4338)$ pentaquark and its partners in the molecular picture}

\author{Mao-Jun Yan}
\affiliation{CAS Key Laboratory of Theoretical Physics, 
  Institute of Theoretical Physics,
  Chinese Academy of Sciences, Beijing 100190}

\author{Fang-Zheng Peng}
\affiliation{School of Physics,  Beihang University, Beijing 100191, China}

\author{Mario {S\'anchez S\'anchez}}
\affiliation{LP2IB (CNRS/IN2P3 – Universit\'e de Bordeaux), 33175 Gradignan cedex,
France}

\author{Manuel {Pavon} Valderrama}\email{mpavon@buaa.edu.cn}
\affiliation{School of Physics,  Beihang University, Beijing 100191, China}

\date{\today}
\begin{abstract}
  The LHCb collaboration has detected a new hidden-charm pentaquark
  with the quantum numbers of a $\Lambda$ baryon:
  the $P_{\psi s}^{\Lambda}(4338)$.
  This pentaquark will be interpreted as a
  $\bar{D}_s \Lambda_c$-$\bar{D} \Xi_c$ resonance within a contact-range theory.
  Here we briefly comment on the relation of the new $P_{\psi s}^{\Lambda}(4338)$
  with the $P^{\Lambda}_{\psi s}(4459)$.
  We find that the $P_{\psi s}^{\Lambda}(4338)$ and $P_{\psi s}^{\Lambda}(4459)$ both
  accept a common description in terms of the same parameters,
  which predicts the existence of a few additional $P_{\psi}^N$,
  $P_{\psi s}^{\Lambda}$, $P_{\psi s}^{\Sigma}$ and $P_{\psi s s}^{\Xi}$
  molecular pentaquarks composed of a charmed antimeson and
  an antitriplet charmed  baryon.
  The most robust of these predicted pentaquarks is a $P_{\psi s}^{\Lambda}$
  with a mass in the $(4235-4255)\,{\rm MeV}$ range, while other two
  interesting ones are a $P_{\psi}^{N}(4150)$ and
  a $P_{\psi}^{\Sigma}(4335)$, the latter basically at the same mass as
  the $P_{\psi}^{\Lambda}(4338)$, with which it might mix
  owing to isospin symmetry breaking effects.
\end{abstract}

\maketitle

{\bf Introduction: }
Recently the LHCb collaboration has announced~\cite{lhcb2022-a,LHCb:2022jad}
the discovery of a new pentaquark with mass and width
\begin{eqnarray}
  M &=& 4338.2 \pm 0.7  \,{\rm MeV} \, , \nonumber \\
  \Gamma &=& 7.0 \pm 1.2\,{\rm MeV} \, ,
\end{eqnarray}
which has been observed in the $J/\psi \Lambda$ mass distribution of
the $B^- \to J/\psi \Lambda \bar{p}$ decay.
It has been named the $P_{\psi s}^{\Lambda}(4338)$ (in the new convention
from~\cite{Gershon:2022xnn}, or $P_{cs}(4338)$ in the previous,
widely-used convention), has the light-quark content of a $\Lambda$ baryon
and its spin-parity is $J^P = \tfrac{1}{2}^-$.
In addition, there seems to be hints of a non-trivial structure in the
$J/\psi \bar{p}$ mass distribution, maybe pointing out towards
a possible $P_{\psi}^{N}$ hidden-charm pentaquark (where
the superscript indicates that its light-quark content
is that of the nucleon).

Here we explore the nature of the new $P_{\psi s}^{\Lambda}(4338)$ pentaquark.
Within the molecular picture it fits well as a heavy-quark
spin symmetry (HQSS) partner of the $P_{\psi s}^{\Lambda}(4459)$~\cite{Aaij:2020gdgold} (often interpreted to be molecular~\cite{Chen:2020uif,Peng:2020hql,Liu:2020hcv,Chen:2020kco}), the reason
being that the $\bar{D} \Xi_c$ and $\bar{D}^* \Xi_c$ diagonal potentials 
are exactly the same (with  corrections originating from coupled
channel effects from the mixing with nearby thresholds~\cite{Peng:2020hql,Chen:2020kco}).
Indeed, a series of previous works have predicted the existence of
a $\bar{D} \Xi_c$ bound state~\cite{Wu:2010jy,Xiao:2019gjd,Wang:2019nvm,Dong:2021juy,Xiao:2021rgp,Yan:2021nio}, with the more recent predictions of
its mass relatively close to where it has been finally
detected, e.g. $(4316-4322)\,{\rm MeV}$ in~\cite{Wang:2019nvm},
$(4329-4337)\,{\rm MeV}$ in~\cite{Dong:2021juy},
$4311\,{\rm MeV}$ in~\cite{Xiao:2021rgp} and
$(4319-4327)\,{\rm MeV}$ in~\cite{Yan:2021nio}.
It should be stressed that this does not necessarily mean that
the $P_{\psi s}^{\Lambda}(4338)$ and $P_{\psi s}^{\Lambda}(4459)$
are molecular, only that they are compatible
with the molecular hypothesis
(see also the recent discussion in~\cite{Karliner:2022erb,Wang:2022tmp};
for non-molecular pentaquark models
check~\cite{Eides:2019tgv,Stancu:2020paw,Ferretti:2020ewe,Ferretti:2021zis}).
Yet, from a Bayesian perspective the convergence of
the previous predictions (which rely on different assumptions:
phenomenological~\cite{Wang:2019nvm,Dong:2021juy,Xiao:2021rgp} or
effective~\cite{Yan:2021nio}) reinforces the belief
that they might be predominantly bound states.

{\bf Effective field theory description:}
We will consider the $P_{\psi s}^{\Lambda}(4338)$ and $P_{\psi s}^{\Lambda}(4459)$
as poles in the $\bar{D} \Xi_c$ and $\bar{D}^* \Xi_c$ two-body
scattering amplitude within a contact-range theory.
With this choice, and if we include
the $\bar{D}_s \Lambda$-$\bar{D} \Xi_c$ coupled channel dynamics,
it will be possible to reproduce the $P_{\psi s}^{\Lambda}(4338)$
as a pole above the $\bar{D} \Xi_c$ threshold.
We will briefly comment on the question of whether these poles can be
interpreted as hadronic molecules at the end of this section,
though we advance that we will only use this nomenclature
for poles below their relevant thresholds.

We describe the two-body interaction within an effective
field theory (EFT) where the low energy degrees of freedom are
the charmed hadrons and the pions (when relevant).
EFTs exploit the existence of a separation of scales in a particular physical
system with the aim of writing scattering amplitudes and observables
as a power series in terms of the ratio $Q/M$, where $Q$ and $M$ are
characteristic low and high energy scales, respectively ($Q \ll M$).
For hadronic molecules $Q$ can be identified with the wave number $\gamma$ of
the bound state, $\gamma = \sqrt{2 \mu B_E}$ with $\mu$ the reduced mass of
the system and $B_E$ its binding energy, while $M$ represents the vector
meson mass or the momentum at which we see the internal structure of
the hadrons. That is, in general we expect $Q \sim (100-200)\,{\rm MeV}$
and $M \sim (0.5-1.0)\,{\rm GeV}$ give or take.
EFT describes hadronic molecules in terms of a non-relativistic
potential composed of a finite- and contact-range piece.
The finite-range piece, which is given by pion exchanges, can be safely
neglected at the lowest order in the expansion (where one pion exchange
is either not allowed owing to HQSS or perturbative and
thus subleading~\cite{Yan:2021nio}).
The contact-range piece, the only one that survives here, represents
the physics at the momentum scale $M$, and we write it down as
\begin{eqnarray}
  \langle \vec{p}\,' | V_C | \vec{p}\, \rangle  = c = \sum_R \lambda_R\,c_R \, ,
\end{eqnarray}
with $\vec{p}$ and $\vec{p}\,'$ the center-of-mass momenta of the initial
and final hadron pair, and where $c$ is a coupling
that can be further subdivided in a sum over $R$ --- the possible
quantum numbers or irreducible representations of
the two-body system under consideration ---
with $\lambda_R$ coefficients.
This potential is singular (it corresponds to a Dirac-delta in r-space),
but can be easily regularized by the introduction
of a cutoff $\Lambda$ and regulator function
\begin{eqnarray}
  \langle \vec{p}\,' | V_C(\Lambda) | \vec{p}\, \rangle = c(\Lambda)\,
  f(\frac{p'}{\Lambda})\,f(\frac{p}{\Lambda}) \, , \label{eq:VC}
\end{eqnarray}
and renormalized by making the coupling dependent on $\Lambda$,
i.e. $c = c(\Lambda)$, and then calibrating this coupling
from the condition of reproducing an observable quantity
(for instance, the binding energy, in which case
it becomes an input of the theory instead of a prediction).
Along this work we will use a Gaussian regulator, $f(x) = e^{-x^2}$,
and a cutoff $\Lambda = 0.75\,{\rm GeV}$, i.e. of the order of
the breakdown scale $M$ of the EFT (in particular in the middle
of the $(0.5-1.0)\,{\rm GeV}$ window we previously
estimated for $M$).
For scattering amplitudes with a well-defined $\Lambda \to \infty$
limit~\footnote{Even if this condition is not met,
  from a more phenomenological perspective a cutoff of the order of
  the $\rho$ meson mass is also a good choice: it maximizes
  the momenta at which the contact-range description is valid
  ($k < \Lambda$) while not being hard enough as to resolve
  the short-range details of the meson-baryon interaction
  (which happens at $\Lambda > m_{\rho}$ if the short-range
  interaction is generated by vector meson exchanges).},
the finite cutoff error $Q/\Lambda$ will be similar to the EFT truncation
error $Q/M$ for $\Lambda \sim M$. 
We will explicitly check whether the cutoff uncertainties are under
control by doubling its value, i.e. we will compare the predictions
of the meson-baryon spectrum for $\Lambda = 0.75\,{\rm GeV}$ and
$1.5\,{\rm GeV}$.

For the calculation of the poles of the two-body scattering amplitude
we use the bound state equation, which for the contact-range
potential of Eq.~(\ref{eq:VC}) can be written as
\begin{eqnarray}
  1 + c(\Lambda)\,\int \frac{d^3 \vec{p}}{(2\pi)^3}\,
  \frac{f^2 (p/\Lambda)}{M_{\rm th} + \frac{p^2}{2\mu} - M_{\rm mol}} = 0 \, ,
  \label{eq:BSE}
\end{eqnarray}
where $M_{\rm mol}$ is the mass of the state we are predicting,
$M_{\rm th}$ the two-body threshold and $\mu$ the reduced mass of
the two-body system.
The evaluation of the loop integral takes the general form
\begin{eqnarray}
  \int \frac{d^3 \vec{p}}{(2\pi)^3}\,
  \frac{f^2(p/\Lambda)}{M_{\rm th} + \frac{p^2}{2\mu} - M_{\rm mol}} =
  -\frac{\mu}{2 \pi}\left[ \gamma +
    \Lambda\,\beta(\frac{\gamma}{\Lambda}) \right] \, ,
\end{eqnarray}
with $\gamma = \sqrt{2\mu\,(M_{\rm th} - M_{\rm mol})}$ the wave number of
the two-body system and $\beta$ a regulator dependent function.
By taking ${\rm{Re}}[\gamma] > 0$ (${\rm{Re}}[\gamma] < 0$)
we will choose Riemann sheet I (II).
Solutions in sheet I (II) correspond to bound (virtual) states.
For the generalization to coupled channels we define
\begin{eqnarray}
    F_{ab} = \delta_{ab} + c_{ab}(\Lambda)\,\int \frac{d^3 \vec{p}}{(2\pi)^3}\,
    \frac{f^2 (p/\Lambda)}{M_{\rm th(b)} + \frac{p^2}{2\mu_b} - M_{\rm mol}} \, ,
  \label{eq:BSE-CC1}
\end{eqnarray}
with the indices $a$, $b$ labeling the channels.
The states (or poles) now correspond to
\begin{eqnarray}
  {\rm det}(F) = 0 \, , \label{eq:BSE-CC2}
\end{eqnarray}
where depending on the combination of Riemann sheets (there are two sheets
per channel: ${\rm{Re}}[\gamma_a] > 0$ and ${\rm{Re}}[\gamma_a] < 0$)
we will talk of bound / virtual states or resonances.

\begin{figure}
  \begin{center}
    \epsfig{figure=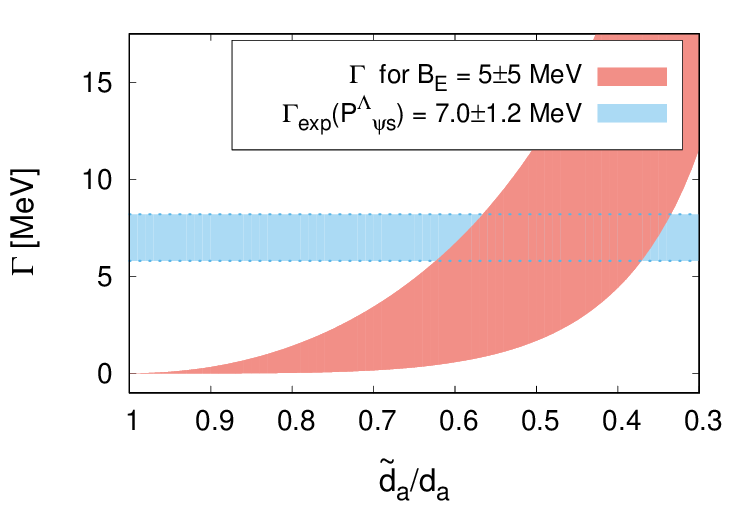,
      width=8.5cm}
    \end{center}
  \caption{
    Partial decay width of a shallow $\bar{D} \Xi_c$ bound state
    into $\bar{D}_s \Lambda_c$ as a function of the ratio of
    the $\tilde{d}_a$ and $d_a$ couplings
    in the contact-range EFT described
    by Eq.~(\ref{eq:Pcs-potential}).
    The binding energy of the state is $B_E = (5 \pm 5)\,{\rm MeV}$ and
    the cutoff is set to $\Lambda = 0.75\,{\rm GeV}$.
    It can be appreciated that the width grows with the difference
    between $\tilde{d}_a$ and $d_a$.
    For comparison, we show the experimental width of
    the $P^{\Lambda}_{\psi s}(4338)$ pentaquark.
  }
\label{fig:width}
\end{figure}

For the complete EFT description of the $\bar{D}^{(*)} \Xi_c$ system
we will refer to~\cite{Peng:2020hql,Yan:2021nio} for further details.
Here we will present an abridged version of Ref.~\cite{Yan:2021nio},
which proposed three possible power countings --- $A$, $B$ and $C$ ---
for pentaquarks containing either antitriplet or
sextet charmed baryons.
Power counting $B$ is restricted to the $\bar{H}_c T_c$ systems,
with $H_c = D$, $D_s$ or $D^*$, $D_s^*$ and $T_c = \Lambda_c$, $\Xi_c$,
and hence well suited for the new $\Lambda$-like pentaquarks.
Within it, the $P_{\psi s}^{\Lambda}(4338/4459)$ is described as a
$\bar{D}_s^{(*)} \Lambda_c$-$\bar{D}^{(*)} \Xi_c$ coupled channel
system with the contact-range potential
\begin{eqnarray}
  V_C(P_{\psi s}^{\Lambda}) =
  \begin{pmatrix}
    \frac{1}{2}(d_a + \tilde{d}_a) & \frac{1}{\sqrt{2}}(d_a - \tilde{d}_a) \\
    \frac{1}{\sqrt{2}}(d_a - \tilde{d}_a) & d_a 
  \end{pmatrix} \, , \label{eq:Pcs-potential}
\end{eqnarray}
with $d_a$ and $\tilde{d}_a$ two independent coupling constants, while
for the other $\bar{H}_c T_c$ systems we have
\begin{eqnarray}
  V_C(P_{\psi}^N / P_{\psi s}^{\Sigma} / P_{\psi ss}^{\Xi} ) = \tilde{d}_a \, .
  \label{eq:Pcs-octet-potential}
\end{eqnarray}
The coupling $d_a$ represents the strength of the diagonal $\bar{D}^{(*)} \Xi_c$
interaction and the $(d_a - \tilde{d}_a)/\sqrt{2}$ linear combination 
determines the partial decay width into the $\bar{D}_s^{(*)} \Lambda_c$
decay channel, which we illustrate in Fig.~\ref{fig:width}
for a shallow $\bar{D} \Xi$ bound state.
Thus, we can determine these two couplings from the mass and width of a
given $\Lambda$-like hidden charm pentaquark,
which is done by solving Eqs.~(\ref{eq:BSE-CC1}) and (\ref{eq:BSE-CC2})
with the $\bar{D}_s^{(*)} \Lambda_c$-$\bar{D}^{(*)} \Xi_c$
potential of Eq.~(\ref{eq:Pcs-potential})
in the (II,I) Riemann sheet.

For the choice of input (i.e. $\Lambda$-like pentaquark), we will either take
the newly discovered $P_{\psi s}^{\Lambda}(4338)$ or the previous
$P_{\psi s}^{\Lambda}(4459)$.
For the latter, there are a single- and a double-peak determination of
its mass~\cite{Aaij:2020gdgold}, with the single-peak mass and width being:
\begin{eqnarray}
  M(P_{\psi s}^{\Lambda}) &=&
  4458.8 \pm 2.9 {}^{+4.7}_{-1.1} \,{\rm MeV} \, , \nonumber \\
  \Gamma(P_{\psi s}^{\Lambda}) &=& 17.3 \pm 6.5 {}^{+8.0}_{-5.7} \,{\rm MeV} \, , 
\end{eqnarray}
and the double-peak:
\begin{eqnarray}
  M(P_{\psi s 1}^{\Lambda}) &=& 4454.9 \pm 2.7 \,{\rm MeV} \, , \nonumber \\
  \Gamma(P_{\psi s 1}^{\Lambda}) &=& 7.5 \pm 9.7 \,{\rm MeV} \, , \\
  \nonumber \\
  M(P_{\psi s 2}^{\Lambda}) &=& 4467.8 \pm 3.7 \,{\rm MeV} \, , \nonumber \\
  \Gamma(P_{\psi s 2}^{\Lambda}) &=& 5.3 \pm 5.3 \,{\rm MeV} \, .
\end{eqnarray}
From HQSS we indeed expect the existence of two $J=\tfrac{1}{2}$, $\tfrac{3}{2}$
$\bar{D}^* \Xi_c$ bound states, as their diagonal potential is exactly
the same (there is no spin dependence).
This degeneracy will be broken by coupled-channel effects, which are naively
expected to be small in comparison with the diagonal interaction.
Reality might be more ambiguous than expectations though, with
Ref.~\cite{Peng:2020hql,Yan:2021nio} finding that
the effect of the nearby $\bar{D}^* \Xi_c'$  and $\bar{D}^* \Xi_c^*$ channels 
ranges between leading order (${\rm LO}$) and next-to-leading order
(${\rm NLO}$) in size.
This means that for a ${\rm LO}$ calculation one might ignore
the aforementioned coupled channel dynamics and still use either of
the previous two poles as input at the price of a slower convergence rate
(the relative ${\rm LO}$ uncertainty with and without
the $\bar{D}^* \Xi_c'$  and $\bar{D}^* \Xi_c^*$ coupled channels
is estimated to be about $0.27$ and $(0.54-0.60)$, respectively,
for the $P^{\Lambda}_{\psi s}(4459)$ as a $\bar{D}^* \Xi_c$
molecule~\cite{Peng:2020hql}).
Alternatively we might average the masses and widths of the two peaks,
which partially cancels out the effects of the coupled channel
dynamics.
What we will do then is to use either the single peak solution, each of
the double peak solutions or the average of the double peak solutions,
i.e.
\begin{eqnarray}
  M(P_{\psi s 12}^{\Lambda}) &=& 4461.4 \pm 2.4\,{\rm MeV} \, , \nonumber \\
  \Gamma(P_{\psi s 12}^{\Lambda}) &=& 6.4 \pm 5.5\,{\rm MeV} \, , 
\end{eqnarray}
for the determination of $d_a$ and $\tilde{d}_a$ from
$P_{\psi s}^{\Lambda}(4459)$, giving us a total of 
five possible determinations.

Here a few comments are in order.
The first is that the $P_{\psi s}^{\Lambda}(4338)$ is close to
the $\bar{D} \Xi_c$ threshold and, as a consequence, its Breit-Wigner mass
and width are likely to differ from that of the pole
in the scattering
amplitude~\cite{Albaladejo:2015lob,Fernandez-Ramirez:2019koa,Yang:2020nrt}
(if it happens to be a composite state).
Indeed, the couplings and spectrum that we will derive
from $P_{\psi s}^{\Lambda}(4338)$ are outliers,
being different than those derived
from the $P_{\psi s}^{\Lambda}(4459)$.

Second, using the width as input implicitly assumes that the decay width
is saturated by the meson-baryon partial width. In this regard,
if we consider the $P_{\psi}^N(4312)$ --- suspected to be a
$\bar{D} \Sigma_c$ bound state~\cite{Chen:2019asm,Chen:2019bip,Liu:2019tjn,Xiao:2019aya,Valderrama:2019chc,Liu:2019zvb,Guo:2019fdo} ---,
the GlueX experiment~\cite{Ali:2019lzf}
found that its branching ratio into $J/\psi N$ is of the order of
the single digit percentages. That is, the $P_{\psi}^N(4312)$ width is
mostly saturated by $\bar{D}^* \Lambda_c$ (except for a small
contribution coming from the width of the $\Sigma_c)$.
Analogously, we will assume the widths of the $P_{\psi s}^{\Lambda}(4338/4459)$
to be saturated by $\bar{D}_s \Lambda_c$ / $\bar{D}_s^* \Lambda_c$.

Third, in its current implementation our calculations are not convergent
for $\Lambda \to \infty$, which eventually manifests as a strong cutoff
dependence for $\Lambda \gg M$.
The reason is the existence of pairs of channels for which the potential
is identical as a consequence of a symmetry but for which
the reduced masses $\mu$ and $\mu'$ are not identical
(e.g. $\bar{D} \Xi_c$ and $\bar{D}^* \Xi_c$ with HQSS).
This eventually generates a divergence proportional to
$\Delta \mu = \mu' - \mu$ for large enough cutoffs
for the state that is being predicted.
Basically, the physical masses are breaking a symmetry of the potential
(e.g. the difference in the reduced masses between $\bar{D} \Xi_c$ and
$\bar{D}^* \Xi_c$, which violates HQSS) and
this manifests as a divergence.
Calculations can be rendered cutoff independent by including $\Delta \mu$
as a subleading correction~\footnote{In this case it will be possible to
  take the $\Lambda \to \infty$ limit and end up with an EFT without
  a cutoff (e.g. the description of the $X(3872)$ as a $\bar{D}^* D$
  molecule found in~\cite{Braaten:2003he}).}$^{,}$
  \footnote{
  This solution only works for symmetries
  connecting {\it both} the potentials and the masses of the particles.
  If the masses are not related by symmetry (but the potentials are),
  as happens with heavy flavor symmetry,
  it is impossible to reformulate the theory
  in a cutoff independent way, as demonstrated in~\cite{Baru:2018qkb}.}.
Yet, predictions from calculations with the physical masses, which are
formally divergent, turn out to have a weak to moderate cutoff
dependence for $\Lambda \sim M$,
see Refs.~\cite{Liu:2019tjn,Valderrama:2019chc,Du:2019pij,Du:2021fmf}
for a few examples involving molecular pentaquarks and cutoffs
ranging from $0.5$ to $1.5\,{\rm GeV}$.
For this reason we follow this later choice (i.e. we use the physical masses),
even though {\it stricto sensu} it is not a genuine EFT calculation
but rather an EFT-inspired calculation.

Fourth, there is the issue of whether the $P_{\psi s}^{\Lambda}(4338)$
can be considered molecular or not:
its relative center-of-mass energy with respect to the $\bar{D} \Xi_c$
threshold is $(1.9 - 3.5\,i)\,{\rm MeV}$ (in the isospin symmetric
limit), where its imaginary part is larger than the real one.
In the semiclassical picture this can be interpreted as the $\bar{D} \Xi_c$
pair dissociating before they orbit each other even once, which does not
comply with our intuitive understanding of a molecule.
For this reason we will not use the term molecular when referring to
the $P_{\psi s}^{\Lambda}(4338)$ and reserve it to relatively
narrow states below a nearby two-body threshold.

\begin{table*}[t]
  \begin{center}
    \begin{tabular}{|ccccccc|}
      \hline \hline
      System & Type &
      Set $B_1$ & Set $B_2$ & Set $B_3$ & Set $B_{4}$ & Set $B_5$\\ 
      \hline \hline
      $\bar{D} \Lambda_c$ & $P_{\psi}^N$ &
      ${(4111.3-4120.2)^V}$ & ${(4153.7-4153.6)^V}$ & $4150.9-4152.0$ & $(4153.7-4153.5)^V$ & $4152.9-4153.4$ \\
      $\bar{D}^* \Lambda_c$ & $P_{\psi}^N$ &
      ${(4256.7-4267.7)^V}$ & $4295.0-4295.0$ & $4291.4-4291.8$ & $4295.0-4295.0$ & $4293.7-4293.9$ \\
      \hline \hline
      $\bar{D}_s \Lambda_c$ & $P_{\psi s}^{\Lambda}$ &
      $4254.8-4254.7$ & $4233.9-4239.4$ & $4237.1-4240.3$ & $4249.5-4251.1$ & $4243.7-4246.1$ \\
      $\bar{D}_s^* \Lambda_c$ & $P_{\psi s}^{\Lambda}$ &
      $4398.4-4398.5$ & $4375.2-4378.5$ & $4378.9-4380.1$ & $4392.1-4392.7$ & $4385.9-4386.8$ \\
      \hline
      \hline
      $\bar{D} \Xi_c$ & $P_{\psi s}^{\Lambda}$ &
      Input & ${4319.0-4321.0}$ & $4315.5-4317.7$ & $4327.8-4329.2$ & $4321.7-4323.6$ \\
      $\bar{D}^* \Xi_c$ & $P_{\psi s}^{\Lambda}$ &
      $4479.2-4478.7$ & Input & Input & Input & Input \\
      \hline
      \hline
      $\bar{D} \Xi_c$ & $P_{\psi s}^{\Sigma}$ &
      ${(4297.4-4308.2)^V}$ & $4336.3-4336.2$ & $4332.8-4333.3$ & $4336.3-4336.3$ & $4335.1-4335.4$ \\
      $\bar{D}^* \Xi_c$ & $P_{\psi s}^\Sigma$ &
      ${(4442.7-4455.3)^V}$ & $4477.5-4477.3$ & $4473.1-4472.6$ & $4477.5-4477.4$ & $4475.8-4475.4$ \\
      \hline
      \hline
      $\bar{D}_s \Xi_c$ & $P_{\psi ss}^{\Xi}$ &
      ${(4401.4-4413.5)^V}$ & $4437.3-4437.2$ & $4433.2-4433.0$ & $4437.3-4437.3$ & $4435.7-4435.6$  \\
      $\bar{D}_s^* \Xi_c$ & $P_{\psi ss}^{\Xi}$ &
      ${(4548.3-4562.3)^V}$ & $4580.9-4580.4$ & $4576.1-4574.7$ & $4581.0-4580.6$ & $4579.0-4578.0$ \\
      \hline \hline
    \end{tabular}
    \caption{Predictions of the $\bar{H}_c T_c$ bound / virtual states
        and resonances (the masses are in units ${\rm MeV}$)
        from the $P_{\psi s}^{\Lambda}(4338/4459)$, depending
        on the specific choice of input.
        The spectrum is calculated for two different cutoffs,
        $\Lambda = 0.75\,{\rm GeV}$ and $1.5\,{\rm GeV}$, where
        the masses are shown in the format
        $M(\Lambda=0.75\,{\rm GeV}) - M(\Lambda=1.5\,{\rm GeV})$.
      Set $B_1$ uses the $P_{\psi s}^{\Lambda}(4338)$ as input
      (i.e. as ``$M_{\rm mol}$'' in Eqs.~(\ref{eq:BSE-CC1}) and
        (\ref{eq:BSE-CC2}) with the potential of Eq.~(\ref{eq:Pcs-potential})),
      leading to the couplings $d_a = -(0.98-0.42)\,{\rm fm}^2$ and
      $\tilde{d}_a = -(0.26-0.25)\,{\rm fm}^2$
      for $\Lambda = (0.75-1.5)\,{\rm GeV}$;
      set $B_2$ uses the $P_{\psi s}^{\Lambda}(4459)$ single-peak solution
      ($d_a = -(1.59-0.54)\,{\rm fm}^2$,
      $\tilde{d}_a = -(0.79-0.39)\,{\rm fm}^2$);
      set $B_3$ uses the lower mass pentaquark ($P_{\psi s1}^{\Lambda}$,
      $M=4454.9\,{\rm MeV}$) of
      the double-peak solution for the  $P_{\psi s}^{\Lambda}(4459)$
      ($d_a = -(1.54-0.54)\,{\rm fm}^2$,
      $\tilde{d}_a = -(1.02-0.44)\,{\rm fm}^2$);
      set $B_4$ uses the higher mass pentaquark ($P_{\psi s2}^{\Lambda}$,
      $M=4467.8\,{\rm MeV}$) of the double-peak solution
      ($d_a = -(1.23-0.48)\,{\rm fm}^2$,
      $\tilde{d}_a = -(0.78-0.38)\,{\rm fm}^2$);
      finally, set $B_5$ uses the average of $P_{\psi s1}^{\Lambda}$ and
      $P_{\psi s2}^{\Lambda}$ as input
      ($d_a = -(1.39-0.51)\,{\rm fm}^2$,
      $\tilde{d}_a = -(0.91-0.41)\,{\rm fm}^2$).
      The masses of the $N$-, $\Sigma$- and $\Xi$-like pentaquarks
      are determined from solving Eq.~(\ref{eq:BSE}) with the potential
      of Eq.~(\ref{eq:Pcs-octet-potential}).
      The superscript $V$ indicates a virtual state.
      For the $\bar{D} \Xi_c$ and $\bar{D}^* \Xi_c$ states we do not show
      the width, as it basically coincides with the width of the input state
      (within half a ${\rm MeV}$).
      Calculations are done in the isospin symmetric limit by averaging
      the masses listed in the Review of Particle
      Physics~\cite{Workman:2022preview}.
    }
    \label{tab:predictions}
  \end{center}
\end{table*}

{\bf Predictions:}
For each of the five determinations considered we predict the spectrum shown
in Table \ref{tab:predictions}: set $B_1$ within the Table indicates
the predictions derived from the $P_{\psi s}^{\Lambda}(4338)$, while
sets $B_2$, $B_3$, $B_4$, $B_5$ instead use the $P_{\psi s}^{\Lambda}(4459)$
(as previously explained) as input.
It is worth commenting that predictions from set $B_1$ are relatively different
from the other sets, which comes as a consequence of the mass and width
of the $P_{\psi s}^{\Lambda}(4338)$ (in particular that it is located
above the $\bar{D} \Xi_c$ threshold).
If the $P^{\Lambda}_{\psi s}(4338)$ turns out to be below this threshold
or is more narrow than its Breit-Wigner determination, it will imply a
larger $\tilde{d}_a / d_a$ ratio (as in Fig.~\ref{fig:width})
and thus predictions more in line to those of
sets $B_2$, $B_3$, $B_4$ and $B_5$.
We also notice that the input dependence of the predictions (the differences
among the aforementioned sets) is larger than their cutoff dependence
within the $\Lambda = (0.75-1.5)\,{\rm GeV}$ range.
In Table \ref{tab:predictions} it can also be appreciated that the farther
a state is predicted from threshold the larger its cutoff dependence is.

The first and most robust prediction is that of the $\bar{D}_s^{(*)} \Lambda_c$
pentaquarks, which bind for all of the sets considered (though,
  depending on the input, predictions vary within a mass range of
  $20\,{\rm MeV}$ and are thus still compatible
  with these two systems not binding).
The $\bar{D}_s \Lambda_c$ ($\bar{D}_s^* \Lambda_c$) partner of
the $P_{\psi s}^{\Lambda}(4338)$ ($P_{\psi s}^{\Lambda}(4459)$)
is located at $(4235-4255)\,{\rm MeV}$
($(4375-4390)\,{\rm MeV}$).
If bound, these pentaquark are expected to be relatively narrow, as their decay
into an anticharmed meson - charmed baryon pair is forbidden.

A second interesting prediction is that of an isovector $\bar{D} \Xi_c$ molecule
with a mass in the $4335\,{\rm MeV}$ region, i.e. a $P_{\psi s}^{\Sigma} (4335)$
pentaquark.
However this state is only there when we use the $P_{\psi s}^{\Lambda}(4459)$
pentaquark as the input for the determination of the parameters
(but not if the $P_{\psi s}^{\Lambda}(4338)$ is used).
From this, it is apparent that the prediction of the $P_{\psi s}^{\Sigma} (4335)$
is more tentative in nature than that of the $\bar{D}_s^{(*)} \Lambda_c$
bound states.
This $P_{\psi s}^{\Sigma}$
has been previously predicted in~\cite{Yan:2021nio} (EFT) or more
recently in~\cite{Wang:2022neq} (QCD sum rules). If molecular,
it should be relative narrow.
Its neutral component $P_{\psi s}^{\Sigma^0}$ will have a small
$J/\psi \Lambda$ decay width owing to isospin breaking effects,
which are easy to quantify.
From SU(3)-flavor symmetry we expect the $\bar{D} \Xi_c(I = 0,1)$ decay
amplitude into $J/\psi \Lambda$ and $J/\psi \Sigma$ to be related
as follows
\begin{eqnarray}
  \langle \bar{D} \Xi_c(I = 0) | H | J/\psi \Lambda \rangle =
  -\frac{1}{\sqrt{3}} \langle \bar{D} \Xi_c(I = 1) | H | J/\psi \Sigma \rangle
  \, ,
\end{eqnarray}
from which it is easy to work out that the isospin breaking branching ratio of
a molecular $P_{\psi s}^{\Sigma^0}$ into these two decay channels will be
\begin{eqnarray}
  \frac{\Gamma(P_{\psi s}^{\Sigma^0} \to J/\psi \Lambda)}{\Gamma(P_{\psi s}^{\Sigma^0} \to J/\psi \Sigma^0)} =
  \frac{1}{3}\,\frac{p_{\Lambda}}{p_{\Sigma}}\,
       {\left| \frac{\Psi_c(0) - \Psi_n(0)}{\Psi_c(0) + \Psi_n(0)} \right|}^2
       \, , \label{eq:branching}
\end{eqnarray}
where $p_{\Lambda / \Sigma}$ refers to the center-of-mass momentum of the final
charmonium - baryon pair, while $\Psi_c({r})$ and $\Psi_n({r})$ are
the charged (${D}^- \Xi_c^+$) and neutral ($\bar{D}^0 \Xi_c^0$)
components of the $P_{\psi s}^{\Sigma^0}$ r-space wave function
evaluated at the origin (${r} = 0$).
For the four determinations of the couplings where there is a $P^{\Sigma}_{\psi s}$
close to threshold, we obtain a branching ratio of
$(0.7-2.4) \cdot 10^{-2}$, $(0.2-0.4) \cdot 10^{-2}$,
$(3.1-6.4) \cdot 10^{-2}$ and
$(0.6-1.6) \cdot 10^{-2}$
for sets $B_2$, $B_3$, $B_4$ and $B_5$, respectively,
and cutoff $\Lambda = (0.75-1.5)\,{\rm GeV}$,
where the closer the prediction to threshold ($B_2$ and $B_4$,
see Table \ref{tab:predictions}) the larger the branching.

From the previous in a first approximation this isovector $\bar{D} \Xi_c$
state is not likely to be easily detectable in the $J/\psi \Lambda$
invariant mass.
Yet, caution is advised.
A closer look at the isospin breaking potential, i.e.
\begin{eqnarray}
    V_C(\bar{D}^{0} \Xi_c^0 - {D}^{-} \Xi_c^+) =
  \begin{pmatrix}
    \frac{1}{2}(d_a + \tilde{d}_a) & -\frac{1}{2}(d_a - \tilde{d}_a) \\
    -\frac{1}{2}(d_a - \tilde{d}_a) &
    \frac{1}{2}(d_a + \tilde{d}_a)
  \end{pmatrix} \, , \nonumber \\
\end{eqnarray}
reveals an interesting pattern: the $(d_a - \tilde{d}_a)$ linear combination
does not only control the meson-baryon decay width of the $P_{\psi s}^{\Lambda}$
but also the size of isospin breaking in $P_{\psi s}^{\Sigma^0}$.
For $d_a = \tilde{d}_a$ we will have twin $\bar{D}^{0} \Xi_c^0$ and
${D}^{-} \Xi_c^+$ bound states with the same binding energy
and an isospin breaking branching ratio of
$p_{\Lambda} / (3 p_{\Sigma}) \simeq 0.5$, see Eq.(\ref{eq:branching}).
This is well above the $\mathcal{O}(10^{-2})$ values we obtained
for sets $B_2$ to $B_5$.
If $|d_a| > |\tilde{d}_a|$ the higher mass ${D}^{-} \Xi_c^+$ state will become
the isovector $P_{\psi s}^{\Sigma^0}$ pentaquark and, as the difference
between the two couplings increases, the branching ratio will
decrease.
That is,
if the mass splitting between the $P_{\psi s}^{\Lambda}$ and $P_{\psi s}^{\Sigma^0}$
is not considerably larger the mass difference between the $\bar{D}^{0} \Xi_c^0$
and ${D}^{-} \Xi_c^+$ thresholds ($2.1\,{\rm MeV}$),
the branching ratio might be sizable.
If this were to be the case, the observed $P_{\psi s}^{\Lambda}(4338)$ peak
might actually be a mixture of a $P_{\psi s}^{\Lambda}$ and
$P_{\psi s}^{\Sigma^0}$ state.

The third prediction that is worth noticing is that of $\bar{D}^{(*)} \Lambda_c$
bound states, which are again only found when using
the $P_{\psi s}^{\Lambda}(4459)$ as input.
In particular, we find a $\bar{D} \Lambda_c$ bound state at about
$4150\,{\rm MeV}$ --- a $P_{\psi}^N(4150)$ pentaquark --- that
might correspond with a possible structure in the $J/\psi p$
invariant mass mentioned in~\cite{lhcb2022-a,LHCb:2022jad}
(though its mass is only reported in~\cite{lhcb2022-a}).
There, when the amplitude contribution from a $P_{\psi}^N$
pentaquark is included, the parameters of this pentaquark happen to be
\begin{eqnarray}
  M(P_{\psi}^N) &=& 4152.3 \pm 2.0 \, {\rm MeV} \, , \nonumber \\
  \Gamma(P_{\psi}^N) &=& 41.8 \pm 6.0 \, {\rm MeV} \, ,
\end{eqnarray}
while the mass and width of the $P_{\psi s}^{\Lambda}$ are left
almost unchanged~\cite{lhcb2022-a}.
Yet, there is a moderate statistical preference for the amplitude model
in which this Breit-Wigner contribution is not present.
Were there to be a $\bar{D} \Lambda_c$ virtual or bound state close to
threshold, the Breit-Wigner parametrization
is unlikely to be an ideal choice.
It is also worth noticing that within the one-boson-exchange model
the $\bar{D}^{(*)} \Lambda_c$ and $\bar{D}^{(*)} \Lambda_c(2595/2625)$
potentials only involve the $\sigma$ and $\omega$ mesons, probably
with similar couplings and cutoffs, i.e.
if one of these systems binds the other is also likely to bind.
In this regard there are previous phenomenological predictions of
$\bar{D} \Lambda_c$~\cite{Chen:2017vai,Shen:2017ayv} and
$\bar{D} \Lambda_c(2595)$~\cite{Burns:2019iih} bound states and arguments
for the interpretation of the $P_{\psi}^N(4457)$ as a $\bar{D} \Lambda_c(2595)$
state~\cite{Burns:2022uiv} (instead of the more usual
$\bar{D}^* \Sigma_c$ identification, which has been argued
not to fully explain the $P_{\psi}^N(4457)$~\cite{Kuang:2020bnk,Burns:2021jlu}).
However, the existence of an unusually long-ranged, $L=1$ one pion exchange
force in the $\bar{D} \Lambda_c(2595)$ system~\cite{Burns:2015dwa,Geng:2017hxc}
has also been shown to make binding much more easy
in this case~\cite{Peng:2020gwk}.
As a consequence, the eventual confirmation of a $\bar{D} \Lambda_c(2595)$
bound state might merely signal the existence of moderate attraction in
$\bar{D} \Lambda_c$, but not necessarily binding.

{\bf Further considerations regarding the $P_{\psi s}^{\Lambda}(4255)$:}
More information might be extracted about the $P_{\psi s}^{\Lambda}(4338)$
and its conjectured lower mass partner from two recent analyses of
the invariant mass distributions in~\cite{Burns:2022uha,Nakamura:2022jpd}.

The amplitude analysis of the $J/\psi \Lambda$ spectrum of
Burns and Swanson~\cite{Burns:2022uha} suggests that there is no
$P_{\psi s}^{\Lambda}(4338)$ in the first place,
but a triangle singularity instead.
It is important to notice that their analysis includes the assumption
that $\tilde{d}_a \geq 0$
though~\footnote{This condition commonly appears in models that saturate
  $\tilde{d}_a$ from vector meson exchange~\cite{Xiao:2019gjd}. Yet,
  this does not take into account the existence of unaccounted attraction
  from other sources (e.g. two-pion exchange).
  For comparison, had we applied this criterion to the neutron-proton
  system,  the result would have been repulsion in the S-wave $S=0,1$
  spin configurations (owing to the strongly repulsive nature of $\omega$
  exchange in this system~\cite{Machleidt:1987hj}),
  in contradiction with experimental evidence.
  We notice that in~\cite{Burns:2022uha} a different linear combinations of
  couplings is used: $A = \tilde{d}_a$ and $\Delta = (d_a - \tilde{d}_a)/2$.
}.
This is incompatible with the determinations we obtain here,
which consistently gets $\tilde{d}_a < 0$.
Had we imposed the condition $\tilde{d}_a \geq 0$, it would have simply been
impossible to reproduce the $P_{\psi s}^{\Lambda}(4338)$ as a resonance
in the (II,I) Riemann sheet.
However, with this constrain we are still able to generate a pole
with the experimental mass of the $P_{\psi s}^{\Lambda}(4338)$
in the (I,II) Riemann sheet --- that is, a sheet that does
not influence physical $\bar{D} \Xi_c$ scattering --- for
$d_a = -(0.62-0.25)\,{\rm fm}^2$ ($\Lambda = (0.75-1.5)\,{\rm GeV}$)
and $\tilde{d}_a = 0$.
This coupling also implies the existence of a $\bar{D}_s \Lambda_c$ virtual
state with mass $M = {(4219.9-4167.3)}^V\,{\rm MeV}$,
which is not expected to be observable either (it is far from threshold and
in the second Riemann sheet).
That is, if we impose the same assumptions as in~\cite{Burns:2022uha}
we also reach the conclusion that the $P_{\psi s}^{\Lambda}(4338)$
is not a $\bar{D} \Xi_c$ scattering pole.

In contrast, the amplitude analysis of Nakamura and Wu~\cite{Nakamura:2022jpd}
explains the $P_{\psi s}^{\Lambda}(4338)$ as a $\bar{D} \Xi_c$
state and predicts the existence of a virtual
$\bar{D}_s \Lambda_c$ pole.
In their analysis the mass and width of the $P_{\psi s}^{\Lambda}(4338)$
are~\cite{Nakamura:2022jpd}
\begin{eqnarray}
  M &=& 4338.0 \pm 1.1 \,{\rm MeV} \, , \\
  \Gamma &=& 1.7 \pm 0.4 \,{\rm MeV} \, ,
\end{eqnarray}
which is narrower than its usual determination from a Breit-Wigner profile.
From the previous mass and width we obtain the couplings
$d_a = -(0.85-0.40)\,{\rm fm}^2$ and $\tilde{d}_a = -(0.25-0.25)\,{\rm fm}^2$
(for $\Lambda = (0.75-1.5)\,{\rm GeV}$),
which at first sight are very similar to the ones obtained
from the Breit-Wigner profile (namely $d_a = -(0.98-0.42)\,{\rm fm}^2$ and
$\tilde{d}_a = -(0.26-0.25)\,{\rm fm}^2$, check the caption
in Table~\ref{tab:predictions}).
There is a difference though when predicting the $\bar{D}_s \Lambda_c$ pole,
which is now a virtual state just below threshold:
\begin{eqnarray}
  M = {(4253.9-4253.4)}^V \, {\rm MeV} \, ,
\end{eqnarray}
which is compatible with the mass extracted in~\cite{Nakamura:2022jpd},
$M = {(4254.6 \pm 0.5)}^V \,{\rm MeV}$.
The results are not expected to be identical, as the contact-range
potential in~\cite{Nakamura:2022jpd} does not follow
the constrains of SU(3)-flavor symmetry
for the non-diagonal terms in Eq.~(\ref{eq:Pcs-potential}).
The possibility that the lower mass partner of the $P_{\psi s}^{\Lambda}(4338)$
is a virtual state is also satisfying from the point of view of
the experimental information available in~\cite{LHCb:2022jad},
where the conjectured $P_{\psi s}^{\Lambda}(4255)$ is not observed.

{\bf Conclusions:}
The $P_{\psi s}^{\Lambda}(4338)$ is a very interesting
pentaquark, which we interpret here as a pole in the $\bar{D} \Xi_c$
  scattering amplitude within a theory where the meson-baryon interaction
  is of a contact-range nature.
On the one hand,
its existence is to be expected from the $P_{\psi s}^{\Lambda}(4459)$:
the $\bar{D} \Xi_c$ and $\bar{D}^* \Xi_c$ S-wave interactions
are identical as a consequence of HQSS, except for perturbations
coming from coupled channel dynamics with nearby thresholds.
Thus, in a first approximation there will be $J=\tfrac{1}{2}$ $\bar{D} \Xi_c$
and $J=\tfrac{1}{2}$, $\tfrac{3}{2}$ $\bar{D}^* \Xi_c$ states
with similar binding energies.
On the other, 
the $\bar{D}^{(*)} \Lambda_c$-$\bar{D}^{(*)} \Xi_c$ coupled channel dynamics
constrains not only the mass, but also the width of
the $P_{\psi s}^{\Lambda}(4338/4459)$ pentaquarks.
This observation allows us to infer more information about the interaction of
the pentaquarks containing an antitriplet
charmed baryon and an anticharmed meson
and predict new states.
The most probable of these predictions is the existence of lower mass,
${\bar{D}}_s^{(*)} \Lambda_c$ partners of the $P_{\psi s}^{\Lambda}(4338/4459)$,
which we denote $P_{\psi s}^{\Lambda}(4250)$ and
$P_{\psi s}^{\Lambda}(4385)$ in reference to
their masses.
Next, though less probable, there might be hints of the existence of
an isovector partner of the $P_{\psi s}^{\Lambda}(4338)$
with a very similar mass --- a $P_{\psi s}^{\Sigma}(4335)$ ---,
also a $\bar{D} \Lambda_c$ pole close to threshold
--- a $P_{\psi}^N(4150)$ ---, and even $\bar{D}_s^{(*)} \Xi_c$ states,
i.e. a $P_{\psi ss}^{\Xi}(4435)$ and $P_{\psi ss}^{\Xi}(4580)$.

Finally, it is worth mentioning that the description of pentaquarks as
meson-baryon systems interacting via contact-range interactions
indeed reproduces the masses of
both the $P_{\psi s}^{\Lambda}(4338/4459)$ and
$P_{\psi s}^{\Lambda}(4459)$ as well as most other pentaquarks, with
the exceptions of the $P_{\psi}^N(4337)$~\cite{Aaij:2021august}
(see e.g.~\cite{Yan:2021nio,Nakamura:2021dix})
and the broad $P_{\psi}^N(4380)$~\cite{Aaij:2015tga} (which we in principle
distinguish from the narrow $P_{\psi}^N(4380)$ predicted
in~\cite{Liu:2019tjn,Xiao:2019aya,Valderrama:2019chc,Liu:2019zvb,Guo:2019fdo}).
This does not necessarily imply that they are meson-baryon states:
for this, a detailed comparison with other competing models will be necessary.
Besides, the methods we use are crude and intended mostly
as a proof of concept:
the most important limitation is the fact that we are using the Breit-Wigner
masses and widths as inputs, which do not necessarily correspond to those
of the poles of the scattering amplitudes, as illustrated by a recent
analysis of the $J/\psi \Lambda$ invariant mass
distribution for the $P_{\psi s}^{\Lambda}(4338)$~\cite{Nakamura:2022jpd}
(though the competing analysis of~\cite{Burns:2022uha} suggest that
this pentaquark might be a triangle singularity).
The eventual extension of this type of analyses to
the $P_{\psi s}^{\Lambda}(4459)$ might potentially confirm or refute
its connection with the $P_{\psi s}^{\Lambda}(4338)$.

\section*{Acknowledgments}

This work is partly supported by the National Natural Science Foundation
of China under Grants No. 11735003, No. 11835015, No. 11975041, No. 12047503
and No. 12125507, the Chinese Academy of Sciences under Grant No. XDB34030000,
the China Postdoctoral Science Foundation under Grant No. 2022M713229,
the Fundamental Research Funds for the Central Universities and
the Thousand Talents Plan for Young Professionals.
M.P.V. would also like to thank the IJCLab of Orsay, where part of
this work has been done, for its long-term hospitality.


%

\end{document}